\documentclass[prl,twocolumn,english,superscriptaddress,floatfix,longbibliography,aps,amsmath,amssymb]{revtex4-2}

\usepackage[utf8]{inputenc}
\usepackage{enumitem}
\usepackage{xcolor}
\usepackage{enumitem}
\usepackage{booktabs}
\usepackage{multirow}
\usepackage{graphicx} 
\usepackage{amsmath}
\usepackage{braket}

\begin{document}
\title{Pump quantization changes optimal measurements for quantum sensing}
\author{S. A. Wadood}
\affiliation{Department of Electrical and Computer Engineering, Princeton University}
\author{L. Bello}
\affiliation{Department of Physics of Complex Systems, Weizmann Institute of Science}
\author{T. Ilias}
\affiliation{Department of Electrical and Computer Engineering, Princeton University}
\author{J. W. Fleischer}
\affiliation{Department of Electrical and Computer Engineering, Princeton University}
    
\date{\today} 

\begin{abstract}
Standard analyses of quantum phase sensing often neglect the overhead of generating squeezed-light by treating the pump as an undepleted classical resource. Realistically, the pump is a quantum object that undergoes significant depletion and entangles with the produced squeezed light, raising the question of how the crossover from a classical to a quantum pump limits the quantum advantage in sensing. \if Surprisingly, a systematic, information-theoretic treatment of phase sensing under pump quantization is missing, a gap we fill here using a quantum Fisher information (QFI)-based analysis. \fi Contrary to prior analyses, we show that quantum advantage persists despite pump quantization, but depends critically on the nonlinear $(\chi^{(2)})$ source dynamics. \if pump quantization does not eliminate quantum advantage. We parametrize the interaction strength, and look at the performance and scaling at different depletion points.\fi \if In particular, we find that at the point of maximum squeezing, the QFI for phase estimation scales as $N^{3/2}$ with the mean pump photon number $N$; this surpasses the shot-noise limit and is saturated by standard photon counting. However, we find that pump quantization precludes squeezing from achieving the Heisenberg limit ($N^2$). \fi
In particular, we demonstrate that Heisenberg scaling emerges in the regime of strong pump depletion, rather than maximum signal squeezing. At maximum squeezing, the quantum Fisher information (QFI) for interferometric sensing of an unknown phase scales as $N^{3/2}$, where $N$ is the pump photon number, surpassing the shot-noise limit but establishing that squeezing itself is insufficient to attain Heisenberg scaling of $N^2$. At maximal pump depletion, where the signal flux is maximized, the QFI scales as $N^2$. In this regime, standard photon counting is suboptimal, while a joint nonlinear measurement of the entire pump-signal system is shown to achieve Heisenberg scaling. Our work shows that pump quantization changes both the optimal operating point and the measurement strategies required to preserve a realistic quantum advantage in phase estimation and, more broadly, many-body quantum dynamics.
\end{abstract}


\maketitle

\label{sec:intro}
\emph{\underline{Introduction}} -- \if Metrology is indispensable for modern optical technologies. Precision measurement not only affects the design and manufacture of optoelectronic components but also improves our understanding of fundamental physical phenomena. These range from measuring gravitational waves \cite{squeezed_states_LIGO} to probing dynamics of atomic systems \cite{Levi_Nick_2015review_on_trapping}.\fi Quantum mechanics not only sets the ultimate limits on the sensitivity of any metrological task, but also offers a path to improve upon classical metrology \cite{giovannetti_quantum_2006,giovannetti_advances_2011}. A hallmark example is the use of squeezed states \cite{caves_quantum-mechanical_1981}, which have been shown to improve measurement of gravitational waves \cite{squeezed_states_LIGO}, sensing beam displacements \cite{Trepshsu2004optimaldisplacementofsingleGaussian}, biomedical imaging \cite{li2024harnessing,kamble2024quantum}, providing resources for continuous variable quantum computing \cite{bourassa2021blueprint}, and even searching for physics beyond the standard model \cite{backes2021quantumaxionsqueezedvacuum}.\par
Nonclassicality is expensive, however, and its generation requires highly efficient and strong nonlinear interactions \cite{mandelandwolf1995book}. Any fair assessment of a claimed quantum advantage needs to factor in these `hidden' costs, and it is debatable whether quantum advantage survives when this overhead is accounted for \cite{caves1994quantumlimitsonbosoniccommunication,giovannetti2004capacity,escher2011general,arvidsson2020quantum,huang2025vast}.\if Although general frameworks have been attempted in the past, the exact answer depends on the task and the resource modeling details.\fi 

 The standard approach to squeezed-state generation~\cite{caves_quantum-mechanical_1981,mandelandwolf1995book} treats the pump as a "free" resource, considering it as a classical drive with no fluctuations. \if which completely neglects the back-action, decoherence, and depletion effects inherent in realistic implementations.\fi  As modern platforms, particularly in nanophotonics \cite{yanagimoto_onset_2022}, achieve stronger nonlinearities and highly efficient squeezing \cite{florez_pump_2020,ding2016quantum}, significant pump depletion is becoming an experimental reality. In these high-flux regimes, the back-action on the pump is non-negligible, the signal and pump states become entangled \cite{xing2023pumpdepletioninparametricamplification,chinni_beyond_2024} and evolve into non-Gaussian states, which fundamentally limits the available squeezing \cite{yanagimoto2022onsetofnongaussiansqueezing}, necessitating a fully quantized treatment. In this work, we consider the following question: does the quantum advantage of squeezing persist in the presence of pump quantization?\par
 Intuitively, pump quantization is expected to introduce  fluctuations~\cite{crouch_limitations_1988}, limiting phase coherence. Several works used phase space heuristics to show that pump quantization suppresses the quantum advantage with squeezing, offering no improvement over using the coherent pump for sensing \cite{kinsler_limits_1993,caves1994quantumlimitsonbosoniccommunication,drobny_quantum_1992,crouch_limitations_1988,hillery1994effect,plimak1994dynamical,hillery1984path}. Nevertheless, to the best of our knowledge, a unified, information-theoretic framework connecting pump quantization, state generation, and metrological performance remains lacking.  

We address this gap by considering the effect of pump quantization and depletion on the sensing capacity of squeezed-state phase interferometry. We find that pump depletion reshapes both metrological sensitivity and optimal measurement strategies. Using an analysis based on the quantum Fisher information (QFI), we show the following:
\begin{enumerate}[label=(\alph*)]
    \item When the phase is encoded in the squeezed signal mode, the QFI for phase sensing cannot beat the classical shot noise limit (SNL) scaling of $N$, where $N$ is the pump photon number.
    \item Mixing the squeezed state with a coherent state and performing Mach-Zehnder interferometry (MZI) yields an $N^{3/2}$ QFI scaling, beating the SNL.
    \item Despite mixing with a coherent state, pump quantization restricts squeezed states from attaining the Heisenberg-limited (HL) scaling of $N^2$.
    \item Operating near the point of \emph{maximum} pump depletion, where the signal flux is maximized, recovers HL scaling of QFI.
    \item Photon counting (or homodyne detection) at the point of maximum depletion is sub-optimal and does not achieve HL scaling. Instead, a nonlinear, time-reversal measurement is shown to saturate the HL \cite{seshadreesan2011parity,anisimov_quantum_2010,nielsen_deterministic_2023,loughlin2025heisenberg,qin2023unconditional,florez_phase_2018}.
\end{enumerate}
(a) is an information-theoretic reformulation of past results based on phase space heuristics \cite{crouch_limitations_1988,kinsler_limits_1993,kinsler_limits_2001}. (b-e) discuss two schemes where the quantum advantage is dynamically reclaimed despite pump quantization. (b) is fundamentally an interference effect between the squeezed signal and the coherent probe \cite{afek_high-noon_2010}. (c) follows directly from (b), showing the limits of squeezed state interferometry. (d) shows that HL scaling can be achieved by tuning dynamics of the nonlinear source generation. Crucially, (e) shows that this transition in source dynamics and information capacity also necessitates a shift in measurement schemes. \if we find that operating near the point of \emph{maximum} pump depletion, where signal flux is maximized, recovers the HL scaling. Crucially, this transition in information capacity also necessitates a shift in measurement schemes. 5) Photon counting (or homodyne detection) at the point of maximum depletion is sub-optimal and does not achieve the Heisenberg bound, and a nonlinear, time-reversal measurement must be implemented to saturate the bound \cite{seshadreesan2011parity,anisimov_quantum_2010,nielsen_deterministic_2023,loughlin2025heisenberg,qin2023unconditional}.\fi \if of This achi  we demonstrate that pump quantization does \textit{not} preclude a quantum advantage; using an analysis based on the quantum Fisher information (QFI), we find that squeezing still yields an $N^{3/2}$ QFI scaling beyond the classical shot noise limit (SNL) scaling of $N$, where $N$ is the pump photon number. However, the quantum nature of pump restricts squeezing from attaining Heisenberg-limited scaling of $N^2$. Counter-intuitively, we find that operating near the point of \emph{maximum} pump depletion, where signal flux is maximized, recovers the HL scaling. Crucially, this transition in information capacity also necessitates a shift in measurement schemes; photon counting (or homodyne detection) at the point of maximum depletion is sub-optimal and does not achieve the Heisenberg bound, and a nonlinear, time-reversal measurement must be implemented to saturate the bound \cite{seshadreesan2011parity,anisimov_quantum_2010,nielsen_deterministic_2023,loughlin2025heisenberg,qin2023unconditional}. \fi We start with a description of our model, followed by an exposition of our numerical results on the information capacity in the presence of pump depletion, as captured by the QFI associated with the phase-encoded state.

\begin{figure}[t]
    \centering
    \hspace*{0.0cm} 
    \includegraphics[trim={0cm 0cm 0cm 0cm},clip,scale=0.75]{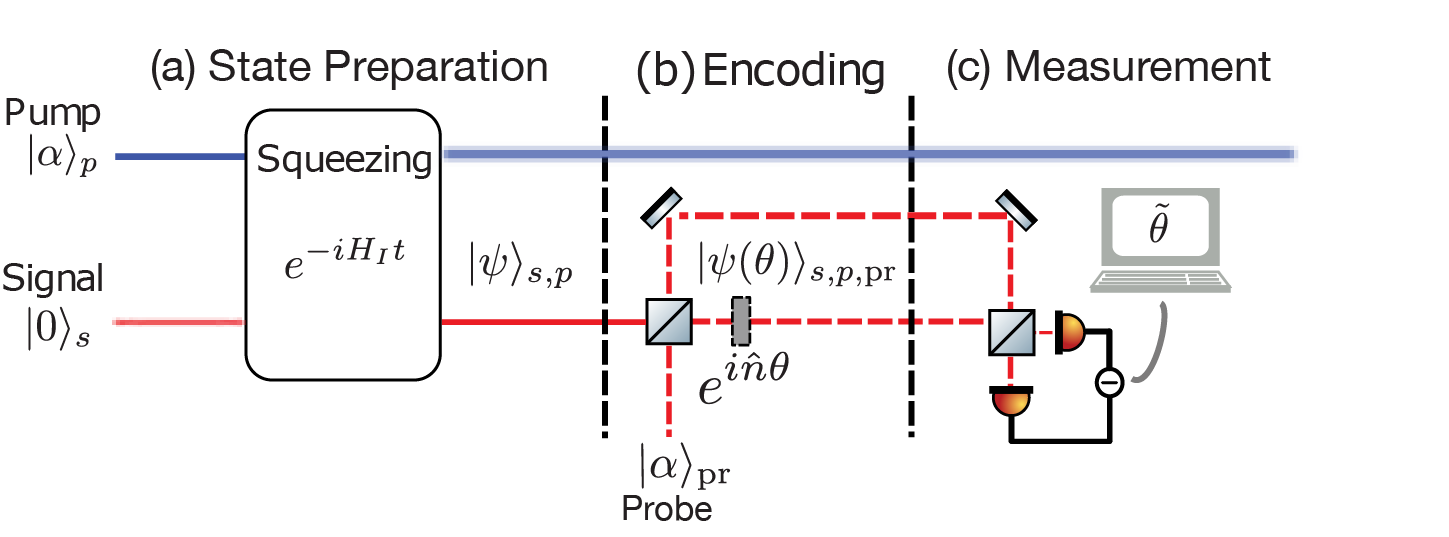}
    \caption{\textbf{(a) Squeezed-state generation}: A pump, in coherent state $|\alpha\rangle_p$, drives a $\chi^{(2)}$ squeezer. The signal is initially in a vacuum state, and interacts with the pump, evolving into a joint signal-pump state $|\Psi\rangle_{s,p}$. \textbf{(b) Phase encoding}: The signal mode is mixed with a local oscillator on the input beam splitter of an MZI \cite{dowlingtakeoka2017fundamentalprecisionofMZI}, and then passed through the phase object. \textbf{(c) Measurement stage}: The photocurrent is post-processed to yield an estimate $\hat{\theta}$.}
    \label{fig:1}
\end{figure}
\begin{figure*}[ht!]
    \centering
    \hspace*{-0.5cm} 
    \includegraphics[trim={0cm 0cm 0cm 0cm},clip,scale=1,width=0.9\linewidth]{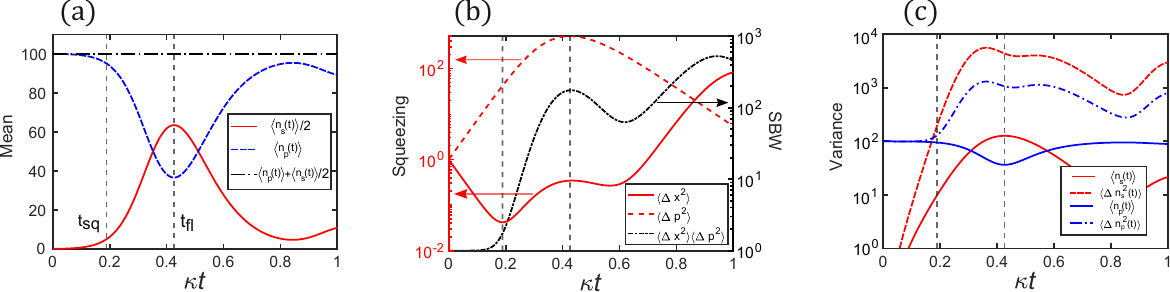}
    \caption{\textbf{Population dynamics in the degenerate parametric oscillator (DPO).} \textbf{(a)} Mean signal and pump photon numbers. The input pump is in a coherent state with mean photon number of $|\alpha|^2 = 100$, and the input signal is in a vacuum state. As a result of energy exchange, the pump is depleted and the signal photon number grows, reaching a peak of just above 60 percent of the pump energy at $t=t_{fl}$, beyond which the energy is transferred back to the pump. For very low depletion values, the signal state has a high fidelity with a squeezed state, with the maximum squeezing occurring at $t=t_{sq}$. In this work, we consider the signal states at $t_{sq}$ and $t_{fl}$ as the input state to be employed in phase estimation. \textbf{(b)} shows the squeezed quadrature variance  ($\langle\Delta x^2\rangle_s$) reaching a minimum at $t=t_{sq}$, while the antisqueezed quadrature ($\langle\Delta p^2\rangle_s$) grows. For $t>t_{sq}$, the signal state is \emph{not} a minimum uncertainty coherent state, and the space-bandwidth product $\braket{\Delta x^2} \braket{\Delta p^2} > 1$. \textbf{(c)} The photon number variances of the signal ($\langle\Delta n^2\rangle_s$) and pump ($\langle\Delta n^2\rangle_p$) grow along with the mean photon numbers. Notably, the signal variance is larger at $t_{fl}$ than at $t_{sq}$. }
    \label{fig:2}
\end{figure*}

\emph{\underline{Squeezed-state Interferometry}} -- Quantum metrology leverages non-classical states to estimate an unknown parameter, $\theta$, with precision surpassing the SNL. In conventional phase sensing, sensitivity is capped by vacuum fluctuations that mix with the probe field at the input of an interferometer~\cite{caves_quantum-mechanical_1981}, as illustrated in Fig.~\ref{fig:1}. Mathematically, we consider a joint input state $\ket{\psi_{in}}_{s,p}=\ket{0}_s\ket{\alpha}_p$ where the signal starts in vacuum, and the pump is prepared in a coherent state $\ket{\alpha}$ with mean photon number $N = |\alpha|^2$. The measured phase is inserted in one arm of the MZI, with the optical path length $\theta$ being imprinted via the interaction Hamiltonian $e^{i\hat{n}\theta}$, $\hat{n}$ being the number operator in the corresponding mode of the MZI arm \if \footnote{Note that nonlinear encodings of the form $\exp[i\hat{n}^k\theta]$ where $k>1$ can dramatically alter the phase sensitivity \cite{boixo2008quantum} and are not considered in this work}\fi. The encoded state $\ket{\psi(\theta)}_{s,p,pr}$, is the state used for phase estimation, and the measurement protocol typically involves photocount detection at the output ports of the MZI [Fig. \ref{fig:1}c]. \par 
Squeezed states are commonly generated through nonlinear optical processes, for example by parametric down-conversion driven by the pump field. The dynamics are modelled by the fully quantized interaction Hamiltonian of a degenerate parametric oscillator~\cite{caves_new_1985,dodonov_nonclassical_2002,walls_squeezed_1983,yuen_two-photon_1976,bello_broadband_2021} (DPO):
\begin{align}
    \hat{H}_I = \frac{i\hbar\kappa}{2}\left(a^{\dagger}_{p}a_{s}^2-a_{p}a_{s}^{\dagger^2}\right),\label{eq:Hamiltonian}
\end{align}
where $\kappa$ denotes the effective single-photon nonlinear coupling rate, while $a_{p},a_{s}$ are the annihilation operators for the pump and signal modes, respectively. To isolate the fundamental physics of pump depletion, we restrict our analysis to a single longitudinal mode for each field and assume perfect phase matching ~\cite{boyd2003nonlinear}. 

In the undepleted, \emph{parametric approximation}, $a_{p}\xrightarrow[]{}\alpha_{p}$ and the resulting Hamiltonian generates an idealized minimum-uncertainty squeezed state which can beat the SNL \cite{slushersqueezedstates1985}, and even achieve HL scaling, by mixing with a coherent probe \if \footnote{We assume that the probe has the same number of photons as the pump. This is consistent with the fact that, experimentally, the probe needs to be phase-locked with the pump, and the pump at $2\omega$ is generated from upconversion of the probe at $\omega$.}\fi in an MZI \cite{seshadreesan2011parity,dowlingtakeoka2017fundamentalprecisionofMZI}. \if  Note that while the pump mode is usually discarded, we include it as part of our analysis.
\fi
\if Squeezed states have been shown to improve measurement of gravitational waves \cite{squeezed_states_LIGO}, sensing beam displacements \cite{Trepshsu2004optimaldisplacementofsingleGaussian}, biomedical imaging \cite{li2024harnessing,kamble2024quantum} and even searching for physics beyond the standard model \cite{backes2021quantumaxionsqueezedvacuum}. Beyond metrology, squeezed states are an essential resource for continuous variable quantum computing \cite{bourassa2021blueprint}. Such states are commonly generated through nonlinear optical processes, for example by parametric down-conversion driven by a strong pump field.
\fi
While the parametric approximation predicts better phase estimation with harder pumping, it completely neglects any pump depletion \cite{florez_pump_2020,ding2016quantum}, decoherence \cite{yanagimoto_onset_2022}, entanglement \cite{xing2023pumpdepletioninparametricamplification,chinni_beyond_2024}, and non-Gaussianity that restricts available squeezing \cite{yanagimoto2022onsetofnongaussiansqueezing}.\par\emph{\underline{Depleted down-conversion}} --  To address these effects, we consider the full quantum Hamiltonian in Eq. (\ref{eq:Hamiltonian}). Since $[n_p+n_s/2,H_I]=0$, the dynamics conserve the total excitations in the system, leading to pump depletion and nonlinear Rabi oscillations between the modes \cite{drobny_quantum_1992}, as shown in (Fig.~\ref{fig:2}a). For an initial state consisting of a coherent pump ($|\alpha|^2 = 100$) and a vacuum signal, the signal photon number $\braket{n}_s$ in the state $\ket{\psi}_{s,p}=\exp{(-iH_It/\hbar)}\ket{\psi_{in}}_{s,p}$ grows until it reaches a peak corresponding to an incomplete transfer of the pump energy at the full depletion time $t_{fl}$. While the parametric approximation predicts a monotonic, exponential decrease in the squeezed quadrature variance $\braket{\Delta x^2}_s$, full quantum dynamics introduce severe phase diffusion as the pump depletes \cite{kinsler_critical_1995,kinsler_limits_2001}, and the quadrature variance is minimized at $t_{sq}$, the point of maximum squeezing. Physically, the evolution $e^{-iH_{I}t/\hbar}$ generates photon-added coherent states in the pump and signal \cite{xing2023pumpdepletioninparametricamplification,chinni_beyond_2024} and fundamentally caps the achievable squeezing \cite{yanagimoto2022onsetofnongaussiansqueezing}. Concurrently, the photon number variances $\braket{\Delta n^2}$ of both the pump and signal grow substantially [Fig. \ref{fig:2}c], maximizing slightly before $t_{fl}$, and not at $t_{sq}$. Appendix A details the numerical methods used in this work. We now address how the DPO dynamics affect phase estimation.  \par
\if Pump back-action fundamentally alters the noise properties of the signal. While the parametric approximation predicts a monotonic, exponential decrease in the squeezed quadrature variance $\braket{\Delta x^2}_s$, full quantum dynamics introduce severe phase diffusion as the pump depletes \cite{kinsler_critical_1995,kinsler_limits_2001}. As shown in Fig.~\ref{fig:2}b, the signal state maintains high fidelity with an ideal coherent state only until $t_{sq}$, the point of maximum squeezing. Approaching $t_{fl}$, the anti-squeezed quadrature uncertainty $\braket{\Delta p^2}_s$ grows rapidly, causing the uncertainty product to exceed unity. This indicates the onset of non-Gaussianity and mixing in the reduced signal state. \fi

\emph{\underline{Sensitivity Analysis}} -- To quantify the phase estimation precision, we utilize the Cramér-Rao bound (CRB), which limits the variance of any unbiased estimator $\tilde{\theta}$ as $\langle\Delta\tilde{\theta}^2\rangle \geq 1/\mathcal{F}(\theta)$. For the pure state $\ket{\psi(\theta)}_{s,p,r}$, the quantum Fisher information (QFI) is defined as $\mathcal{F}(\theta) = 2\text{Tr}[(\partial_\theta \rho)^2]$, representing the maximum extractable information about $\theta$ independent of the specific measurement scheme. For the MZI in Fig.~\ref{fig:1}b, the QFI is determined by the photon number fluctuations in the lower arm, $\mathcal{F} = 4\langle(\Delta\hat{n}_1)^2\rangle$ \cite{lang_optimal_2013, dowlingtakeoka2017fundamentalprecisionofMZI}~\footnote{Appendix D discusses alternative phase generators such as $\hat{n}_1 - \hat{n}_2$.} \if may shift the absolute QFI, but do not alter the underlying scaling behavior \cite{dowlingtakeoka2017fundamentalprecisionofMZI}.\fi. Assuming a signal-probe product state at the MZI input and noting that odd-order signal moments vanish when $\ket{\psi_{in}}_{s,p}$ evolves under $H_{I}$, the QFI evaluates to:
\begin{align}
  \mathcal{F}&=2|\alpha|^2+\braket{\Delta
  n_s^2}+\braket{n_s}+2|\alpha|^2\braket{\Delta a^{\dagger}_s \Delta
  a_s} \nonumber\\
  &-2\Re{|\alpha|^2e^{-i2\phi}\braket{\Delta a^{2}_s}},
  \label{eq:QFI}
\end{align}
where $|\alpha|^2$ and $\phi$ represent the probe intensity and phase, respectively, and $\Delta a=a-\braket{a}$.

\par
\begin{figure}[t]
\centering
\hspace*{-0.3cm} 
\includegraphics[trim={0cm 0cm 0cm 0cm},clip,scale=1]{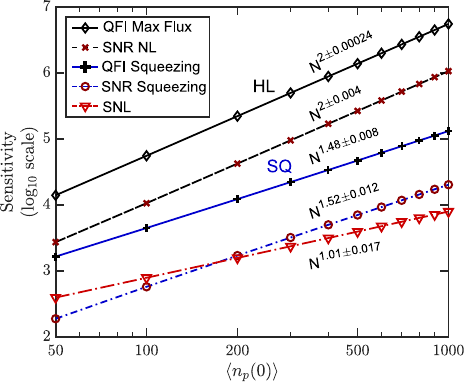}
\caption{\textbf{QFI Scaling for the DPO.} For the DPO biased at maximum flux ($t_{fl}$), the QFI (solid black line and black diamonds) shows Heisenberg-limited (HL) scaling. For the DPO biased at maximum squeezing ($t_{sq}$), the QFI (solid blue line and black plus signs) does not attain the HL, but instead scales as $N^{1.48\pm0.008}$, still beating the shot noise limit (SNL). The sensitivities of the respective measurement schemes are shown in dashed lines. For the nonlinear measurement in Fig. (\ref{fig:4}) (dashed black lines and brown crosses), the sensitivity shows HL scaling, while the Mach-Zehnder scheme shown in Fig. (1) achieves similar scaling as the QFI at $t_{sq}$. For $\alpha=0,~t=t_{sq}$, squeezing achieves the SNL (red dashed line and red triangles). Note that the dashed and solid lines are formed by connecting the data points computed numerically, while the scaling shown is obtained from a linear fit to the numerical data. The prefactors to the scaling are detailed in Appendix C. }
\label{fig:3}
\end{figure}

 Equation (\ref{eq:QFI}) provides an intuitive framework for quantifying the quantum advantage for any arbitrary signal state. We analyze the scaling of $\mathcal{F}$ versus the total photon resource $N$. When the DPO is biased at $t_{sq}$, we find that $\braket{\Delta
  n_s^2}$ scales as $\sim N$ and $\braket{n_s}$ scales as $\sim\sqrt{N}$; results (a-c) in the introduction follow immediately. (a) When the probe is in vacuum ($\alpha=0$), $\mathcal{F}=\braket{\Delta
  n_s^2}+\braket{n_s}$ and we find a scaling of $N^{1.01\pm0.017}$ which is just the semiclassical SNL dictated by the coherent probe. Without the probe, squeezing itself cannot provide any quantum advantage, as also found by phase space arguments in \cite{kinsler_limits_1993,crouch_limitations_1988} and explained in Appendix B \if\footnote{Notice that for $\kappa=0$, that is, when the signal is in vacuum, $\mathcal{F}=2|\alpha|^2=2N$, which is just the semiclassical SNL dictated by the coherent probe.}\fi. (b) However, when mixed with a probe with mean photon number $N\approx |\alpha|^2$, the QFI scales as $N^{1.48\pm0.008}$, as shown in Fig. (\ref{fig:3}). The advantage stems from the interference term $2|\alpha|^2\braket{\Delta a^{\dagger}_s \Delta
  a_s}$ in Eq. (\ref{eq:QFI}) which is expected to scale as $\sim N^{3/2}$. Interference between the probe and squeezed state at the beamsplitter \cite{caves_quantum-mechanical_1981,afek_high-noon_2010} preserves the quantum advantage despite pump quantization. (c) Finally, we see that no term scales as $N^2$ in Eq. (\ref{eq:QFI}); squeezing itself does not achieve Heisenberg scaling.\par 
  Error propagation analysis shows that photon counting at the MZI output achieves this near-$N^{3/2}$ scaling. For an MZI phase-biased near $\theta=\pi/2$ and $\phi=\pm\pi/2$, linearized error propagation yields the SNR:
\begin{align}
  S_{PC}=\frac{\left(\partial_{\theta}\braket{n_{D}}\right)^2}{\braket{\Delta n^2_{D}}}=\frac{(|\alpha|^2-\braket{n_s})^2}{\braket{n_s}+|\alpha|^2\braket{(\Delta x)^2}}
  \label{eq:SNR_photon_counting}
\end{align}
where $x=i(a^\dagger_s-a_s)$ is the squeezed quadrature and $n_{D}$ is the photon number difference in the two output ports of the MZI. Our numerically computed $S_{PC}$ scales as $N^{1.52\pm0.012}$ (Fig.~\ref{fig:3}), successfully beating the SNL, as also expected from the $\sqrt{N}$ scaling of $\braket{n_s}$ at $t_{sq}$ \footnote{We attribute the minor discrepancy to numerical errors in calculating the matrix exponential and to the finite size of the pump photon number distribution that we are sampling. \if technically speaking, we are in the mesoscopic regime, and cannot probe the asymptotic case of infinitely high photon numbers without further approximations to the dynamics.\fi Nevertheless, the numerical value of the photon counting SNR remains bounded by the QFI ($S_{PC} \leq \mathcal{F}$), as expected.}.

\emph{\underline{Optimal Operation}} -- 
\if Until now, we have shown that squeezing with photon counting can beat the SNL, even in the presence of pump quantization. This scheme however does not alone suffice to achieve Heisenberg-limited scaling. \fi 
A natural question is whether HL can be achieved with pump quantization, and if so, what sets the optimal operating point of the squeezer. We therefore consider dynamics beyond $t_{sq}$, where the signal evolves beyond the coherent-state manifold. A key feature is a maximum in the signal flux at $t=t_{fl}$, when most of the pump energy has been transferred and $\langle n_s \rangle \propto N$. Furthermore, the photon number variance is maximized slightly before $t_{fl}$, not at $t_{sq}$, and operating at the optimal squeezing value actually \emph{limits} the achievable sensitivity. This indicates that it is optimal to bias the DPO for maximal flux, interacting for time $t_{fl}$. By properly accounting for the quantum state of the pump in the measurement state, this highly non-Gaussian, depleted operating point can be leveraged to achieve higher sensitivity.\par
We find that \textit{$\mathcal{F}$ exhibits HL scaling of $N^{2\pm0.00024}$ if the squeezer is biased at $t=t_{fl}$}. Note that two terms in Eq. (\ref{eq:QFI}) exhibit the quadratic scaling at $t_{fl}$, namely, the signal variance $\braket{\Delta n_s^2}$ and the cross term between the signal and probe mean $2\alpha^2\left(\braket{\Delta a^{\dagger}_s \Delta a_s}\right)$. That the signal variance scales as $N^2$ indicates that the signal photon number distribution spans the full support of the Hilbert space associated with each pump Fock state, as explored in Appendix E. \par

Maximizing the flux is an intuitive, first-order, and phase-insensitive design principle that engineers can use to make efficient nonlinear devices in highly non-Gaussian regimes. This could amount to optimizing the length of a nonlinear crystal or an (adiabatic) pump pulse duration. As Fig. (\ref{fig:2}) shows, optimizing flux usually conflicts with, and is often computationally simpler than, optimizing squeezing. Our results demonstrate an optimum of flux and squeezing that results in a higher QFI for phase estimation and pave the way for future experiments to incorporate such design principles.\par

\begin{figure}[t]
    \centering
    \hspace*{-0.35cm} 
    \includegraphics[trim={0cm 0cm 0cm 0cm},clip,scale=1.1]{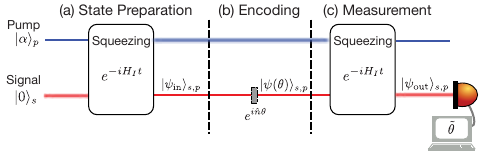}
    \caption{\textbf{The SU(1,1) measurement scheme}: The phase is encoded in the signal beam that passes through the phase element. The pump and signal are then passed through another DPO with the same interaction strength as the first DPO followed by a photon number measurement in the signal mode. For a phase biased at $\theta=\pi/2$, the sensitivity of this scheme shows Heisenberg scaling.}
    \label{fig:4}
\end{figure}

While the QFI exhibits HL scaling by a simple change in the operating point of the squeezer, it comes at the expense of a more sophisticated measurement requirement to saturate the QFI. As we show in Appendix C, photon counting only achieves shot noise scaling if the squeezer is flux-optimized. Intuitively, this means that photon counting is not utilizing the complete information available in the photon statistics of the flux-optimized state, since Eq. (\ref{eq:SNR_photon_counting}) is only sensitive to the signal mean and its quadrature fluctuations. To this end, one would expect an improved SNR scaling with a measurement scheme that is sensitive to higher order photon number moments. Different schemes have been proposed in the squeezing literature, such as parity detection \cite{seshadreesan2011parity,anisimov_quantum_2010}, a generalized homodyne scheme measuring $\braket{x^2}$ instead of $\braket{x}$ \cite{nielsen_deterministic_2023,loughlin2025heisenberg}, or SU (1,1) interferometers using time-reversal schemes \cite{dowlingtakeoka2017fundamentalprecisionofMZI,qin2023unconditional,linnemann2016quantum,kamble2024quantum,florez_phase_2018}. While a complete review of such schemes is beyond the scope of this paper, a recurring theme in these works is the use of nonlinear (in the photon number) schemes to exhaust the information available in the non-Gaussian, nonclassical states of the kind generated in the squeezer when it is flux-optimized. A typical implementation of such a nonlinear measurement is shown in Fig. \ref{fig:4}, where both the encoding and measurement schemes are now different from that in Fig. \ref{fig:1} \footnote{A similar setup was shown for non-degenerate downconversion in \cite{florez_phase_2018}, although a unique optimal operating point, the relation to squeezing, and the associated QFI was not explored.}. The phase is encoded directly in the signal beam, while the measurement involves another evolution under $H_I$ with the same strength and interaction time used in state preparation. Note that this is a joint measurement with the signal and pump beams, and we finally measure the photon number in the signal beam. The QFI, $\mathcal{F}_{NL}=4\braket{\Delta n^2_s}$ and the SNR, $S_{NL}$, of this nonlinear scheme is defined as $\left(\partial_{\theta}\braket{n_{s}}\right)^2/\braket{\Delta n^2_{s}}$. When biased at the dark fringe at $\theta=\pi/2$, $S_{NL}$ achieves HL scaling, as shown in Fig. (\ref{fig:3}). The nonlinear scheme is not always optimal; when operated at $t_{sq}$ (squeezing-optimized) instead of $t_{fl}$ (flux-optimized), both $\mathcal{F}_{NL},S_{NL}$ exhibit shot noise scaling (see Appendix C for details). Thus, neither photon counting nor nonlinear detection provides a quantum advantage for all interaction strengths. The optimality of a detection scheme is highly sensitive to the encoded quantum state, a well-established feature in quantum-enhanced metrology that is further amplified by the effects of pump quantization and depletion.

\if
Simple, closed form expressions do not exist for the signal mode dynamics in case of a quantized pump, and one often has to resort to numerical methods.  It is not just that the pump is depleted, the pump becomes entangled with the signal \cite{chinni_beyond_2024,yanagimoto_onset_2022}, and the signal is not a Gaussian state anymore. \fi

\label{sec:conclusions}
\emph{\underline{Discussion and Conclusions}} -- In summary, we have established a framework for quantum sensing that accounts for the finite resources and dynamical back-action on the generating pump. While previous analyses suggest that squeezing cannot beat the SNL and is useful only when high pump power is not feasible such as in the case of phototoxicity \cite{li2024harnessing}, radiation pressure \cite{caves_quantum-mechanical_1981} or thermal lensing \cite{zhao2006compensation},  we demonstrate that the genuine quantum enhancement in sensitivity remains robust despite pump back-action. Specifically, squeezing can outperform SNL even with pump quantization, achieving a QFI scaling of ${\sim}N^{3/2}$. Nevertheless, pump quantization precludes squeezing from achieving HL scaling, which is only approached deep into the depletion regime where the mode is manifestly unsqueezed, and traditional linear detection is suboptimal. Instead, a nonlinear scheme jointly measuring the pump and signal, such as an SU(1,1) interferometer \cite{shaked_lifting_2018, yurke_su2_1986}, is optimal in the depleted regime. \par 
Our single-mode model can be readily extended to capture the multimodal, open system physics of realistic systems. A critical avenue is to extend our fully quantized 3-body framework to encompass the 4-body $\chi^{(3)}$ dynamics which generate squeezed states in nanophotonic waveguides, optical fibers, and atomic ensembles \cite{boyd2003nonlinear}. An open question is whether the robust ${\sim}N^{3/2}$ sensitivity scaling observed here persists or if higher-order nonlinear processes or losses enforce stricter practical bounds on the quantum advantage. Furthermore, our analysis is qualified by a local, as opposed to a global, sensing protocol \cite{dowlingdurkin2007local}, and saturating multi-parameter Fisher information bounds (vs. single parameter in this work) requires judicious estimator designs, priors determined by experimental details, and taming finite data sizes. Such details are germane to any practical application of quantum metrology \cite{hradil2026realistic} and offer productive routes to extend our analysis.\par

By treating the pump as a quantized degree of freedom, we have shown that the fundamental limits of phase estimation are dictated not merely by the properties of the generated squeezed state, but by the joint coherence and entanglement of the entire pump-signal system. The framework, based on the 3-body DPO Hamiltonian of Eq. (\ref{eq:Hamiltonian}), informs not only quantum optical sensing but also a diverse physical landscape including the $N$-atom Tavis-Cummings model \cite{tavis1969approximate,mohamadian2019tavis}, Brillouin scattering \cite{enzian2018observation}, phonon lasing \cite{zhangnickvamivakas2026twomodesqueezing}, circuit QED \cite{ashhab2010qubit,abdo2013nondegenerate}, nonlinear gravitational waves \cite{sreenath2026squeezed}, and quantum thermodynamics \cite{maslennikov2019quantum}. These results reconcile the idealized notions of a ``probe state'' generated by quantum dynamics across these different systems with the constraints of the probe's physical generation. To achieve any nonclassical advantage, the design of the sensor must encompass the full cycle of the quantum resource---from its creation by a quantized pump to its final readout in a correlated, nonlinear measurement.\par
Beyond sensing, our work informs on the deeply related question of the impact of quantization of a classical control signal on quantum dynamics \cite{senitzky1996can,kimblevan2001classicalcharacterofcontrolfields,twyeffort2022defining}. More broadly, we demonstrate that quantizing traditionally semiclassical resources is not merely a bookkeeping exercise for accurate performance bounds, but a fundamental prerequisite for designing optimal quantum protocols.
\section{*Acknowledgements}
S. A. W. and L. B. acknowledge useful discussions with L. P. McGuinness that triggered this work. S. A. W. acknowledges OpenAI's LLM  for pointing to related work in \cite{shchesnovich2008fock,kirkby2019quantum}.\par

L. B. and T. I. contributed equally to this work.

\bibliography{references}

\appendix
\section*{Appendix A: Numerical Methods}
In this section, we describe the numerical simulation methods used to obtain the results presented in the main text, closely following approaches employed in previous works \cite{chinni_beyond_2024,florez_phase_2018,drobny_quantum_1992}.\par
To simulate the dynamics, we expand the input state $\ket{\psi_{in}(t=0)}_{s,p}=\ket{0}_s\ket{\alpha}_p$ in a joint Fock basis as $\ket{0}_s\ket{\alpha}_p=\sum_{n}c_{n}\ket{0}_s\ket{n}_p$, where $|c_n|^2$ is Poisson distributed with mean $N=|\alpha|^2$. \if $ \ket{\psi_{in}}_{s,p}=\sum_{n_s,n_p}c_{n_s,n_p}\ket{n_s}_s\ket{n_p}_p$.\fi Within the rotating-wave approximation, the DPO Hamiltonian in Eq. (\ref{eq:Hamiltonian}) is time-independent, and thus direct integration of the Schr\"odinger equation yields the time-evolved state $\ket{\psi_{in}(t)}_{s,p}=\exp{[-iH_{I}t/\hbar]}\ket{\psi_{in}(t=0)}_{s,p}$. To compute the matrix exponential directly, we can use the matrix elements of the Hamiltonian in the joint Fock basis $\ket{n_s}_s\ket{n_p}_p$. A more efficient approach to evolve the input state $\ket{\psi_{in}}_{s,p}$ is to use the fact that the Hamiltonian commutes with the total photon number operator $\hat{M}=n_p+n_s/2$, i.e., $[\hat{M},H_I]=0$, and thus a shared set of eigenvectors exists between $H_I$ and $\hat{M}$. Consequently, the evolution of a state with a given eigenvalue $M$ does not couple to states with different total photon number $M'$. Alternatively, the evolution of states with different pump photon numbers in the initial state $\ket{\psi_{in}}_{s,p}$  is decoupled, and we can evolve the subspace corresponding to each $M$ independently. Denoting $\ket{n}_M\equiv\ket{2(M-n)}_s\ket{n}_p$ as a ket in the subspace containing $M$ photons shared between the signal and pump mode, we obtain the tridiagonal matrix
{\small
\begin{align}
   &_M\bra{n}H_I\ket{n'}_M=\begin{cases}
      -ia_{nn'},&\textrm{for~ $ n'=n+1$};\\
      ia_{nn'},&\textrm{for~ $ n'=n-1$};\\
      0 &\textrm{~otherwise}
   \end{cases},\label{eq:matrix_representation_Hamiltonian}
\end{align}
}where $a_{nn'}=(\hbar\kappa/2)\sqrt{n(2M-2n+1)(2M-2n+2)}$ and $0\leq n,n'\leq M$.\par
For the simulation, $M$ is varied from $M_{min}=N-5\sqrt{N}$ to $M_{max}=N+5\sqrt{N}$, which covers the tails of the Poisson distribution that are at least 5 orders of magnitude smaller than the peak value. For each $M$, the $t=0$ ket state $\ket{n}_M$ is represented by a column vector whose $n^{th}$ element is equal to unity and the rest are zeros. At $t=0$, $\ket{n=M}_M$, since all the photons are in the pump mode, and the ket is given by $[\overbrace{0,...,0,}^{\text{M-1 entries}}~1]^{T}$.  We directly exponentiate the $M-$dimensional matrix obtained through Eq. (\ref{eq:matrix_representation_Hamiltonian}), and use the MATLAB function  \textit{expmv\_tspan} \cite{al-mohy_higham_expmv_2010} developed in \cite{higham_al2011computing} to compute the time-evolved ket. For the simulation, $\kappa=1$, and the temporal axis is set to $t=[0.001:0.0001:1]$ to efficiently sample the dynamics over different pump powers. We simulated the interaction for $N={[50,100:100:1000]}$.\par
The time-evolved state can then be used to compute the physical observables such as $\braket{n_s},\braket{n_p},\braket{x^2}_s,\braket{p^2}_s,\braket{\Delta n_s^2}$ which are plotted in Fig. (\ref{fig:2}) of the main text and are used in computing the QFI and SNR plotted in Fig. (\ref{fig:3}) of the main text. \if [CAN END APPENDIX HERE] We plot the signal and pump mean for longer times in Fig. (SA.1). 
While the distribution spans the infinite dimensional Hilbert space, we restrict the sum over $n$ to $N\pm5\sqrt{N}$,  
With the highest dimension of the pump Hilbert space equal to $M_{max}=N+\sqrt{5}$, the ket $\ket{n\leq M_{max}}_M$ is represented as a column vector of dimension $M_{max}$
{\small
\begin{align}
&_s\bra{n'_s}_p\bra{n'_p}H_I\ket{n_s}_s\ket{n_p}_p=\frac{i\hbar\kappa}{2}\big(_s\bra{n'_s}_p\bra{n'_p}a^{\dagger}_{p}a_{s}^2\ket{n_s}_s\ket{n_p}_p\nonumber\\&-\bra{n'_s}_p\bra{n'_p}a_{p}a^{\dagger^2}_{s}\ket{n_s}_s\ket{n_p}_p\big)\nonumber\\
&=\frac{i\hbar\kappa}{2}\big(\sqrt{(n_p+1)(n_s)(n_s-1)}\delta_{n'_p,n_p+1}\delta_{n'_s,n_s-2}\nonumber\\
    &-\sqrt{(n_p)(n_s+1)(n_s+2)}\delta_{n'_p,n_p-1}\delta_{n'_s,n_s+2}\big).
\end{align}
}  
Plugging this state in the Schrodinger equation with the DPO Hamiltonian in Eq. (\ref{eq:Hamiltonian}) of the main text, we get
{\small

\begin{align}
    i\hbar\partial_t\ket{\psi_{in}}_{s,p}&=H_{I}\ket{\psi_{in}}_{s,p},\nonumber\\
     i\hbar\sum_{n_s,n_p}\dot{c}_{n_s,n_p}\ket{n_s}_s\ket{n_p}_p &= \sum_{n_s,n_p} \frac{i\hbar\kappa}{2}\left(a^{\dagger}_{p}a_{s}^2-a_{p}a_{s}^{\dagger^2}\right) c_{n_s,n_p}\ket{n_s}_s\ket{n_p}_p
\end{align} 
}
Taking inner product on the left with $\ket{n'_s}_s\ket{n'_p}$, we get
{\small
\begin{align}
    \dot{c}_{n_s,n_p}&=\sum_{n_s,n_p}\frac{\kappa}{2}\big(\sqrt{(n_p+1)(n_s)(n_s-1)}\delta_{n'_p,n_p+1}\delta_{n'_s,n_s-2}\nonumber\\
    &-\sqrt{(n_p)(n_s+1)(n_s+2)}\delta_{n'_p,n_p-1}\delta_{n'_s,n_s+2}\big)c_{n_s,n_p}\ket{n_s}_s\ket{n_p}_p,\nonumber\\
    \dot{c}_{n_s,n_p}&=\frac{\kappa}{2}\big(\sqrt{(n_p)(n_s+2)(n_s+1)}c_{n_s+2,n_p-1}-\sqrt{(n_p+1)(n_s)(n_s-1)}c_{n_s-2,n_p+1}\big)
\end{align}
}
 Because the DPO Hamiltonian in Eq. (\ref{eq:Hamiltonian}) conserves total the photon number, each pump Fock state evolves as an independent subspace.  The dynamics of the squeezer are shown in Fig. (\ref{fig:2}) for $N=100$. For short times, the pump flux ($\braket{n_p(t)}$) is nearly undepleted and the signal flux ($\braket{n_s(t)}$) is very small, as shown in Fig. (\ref{fig:1} a). For longer times, a significant fraction of the pump is depleted, and the signal flux is high. The dynamics for longer time, detailed in Appendix A,  show oscillation of flux between the signal and pump, with eventually quasi-steady states achieved for coherent pump [REFERENCE other papers quesada etc]. For squeezing experiments, one would like to minimize the squeezed quadrature variance $\braket{\Delta x^2}_s$. Figure (\ref{fig:2}b) shows that the quadrature variance initially decreases till it reaches a minimum at $t=t_{sq}$, after which the variance starts to increase. We also note that the signal mode does not remain a miminum-uncertainty state at $t>t_{sq}$. For optimal squeezing, we would therefore `bias' the squeezing interaction duration such that the interaction stops at $t=t_{sq}$. This would amount to, for example, optimizing the length of a nonlinear crystal or an (adiabatic (because we are doing single mode)) pump pulse duration.\par
\fi
\section*{Appendix B: Previous work on phase estimation with quantized pump}
In this section we compare our results with past work on limits to squeezing-based phase estimation in the presence of pump depletion \cite{kinsler_limits_1993,kinsler_limits_2001}, pump phase fluctuations in the parametric approximation \cite{crouch_limitations_1988,hillery1984path}, and energy balance arguments \cite{paris_small_1995,barnett2003ultimate}.\par
\emph{Phase space SNR --} From a phase space perspective, a physically motivated SNR for squeezed state phase estimation can be defined as the ratio of i) mean photon number $\braket{n_s}$ in the squeezed state (signal) and ii) the variance in the squeezed quadrature $\braket{(\Delta x)^2}$ (noise). For a coherent state $\ket{\alpha}$, for example, $\braket{n_s}=|\alpha|^2$, and the variance $\braket{\Delta x^2}=1$ of the quadrature $x=i(a^\dagger-a)$. The SNR is then given by $|\alpha|^2/\braket{\Delta x^2}=|\alpha|^2=N$, which is the SNL. For a squeezed state, it was shown in \cite{crouch_limitations_1988,kinsler_limits_2001,hillery1984path} that $\braket{(\Delta x)^2},\braket{n_s}$ respectively scale as $N^{-0.5},N^{+0.5}$ \textit{when the DPO is biased at maximum squeezing (or minimum quadrature variance)}. Thus, the phase space SNR for a squeezed state $\braket{n_s}/\braket{(\Delta x)^2}$ scales as $N$, equivalent to the SNL. Note that the phase space SNR provides a lower bound for the variance $\braket{\Delta \theta^2}$; to compare with the previous literature using the standard deviation $\Delta \theta=\sqrt{\braket{\Delta \theta^2}}$, one should use the square root of the SNR defined above. This can be interpreted as a direct consequence of pump quantization: the dynamical phase fluctuations of the pump limit the available squeezing, and the finite pump energy limits the generated flux. This result was interpreted to suggest that the phase resolution of any interferometric scheme employing squeezed vacuum cannot surpass the SNL, and that the advantage of squeezing lies primarily in reduced power requirements rather than in improved scaling.\par 
Such phase space metrics are based on the intrinsic parameters of the generated state and do not consider either the phase encoding schemes or measurement statistics.  These results can be recovered in our formalism by setting $\alpha=0$ in Eq. (\ref{eq:QFI}) to obtain $\mathcal{F}=\braket{n_s}+\braket{\Delta n^2_s}$. At the point of maximum squeezing, $\braket{n_s},\braket{\Delta n^2_s}$ scale as $N^{0.5},N$ respectively, and $\mathcal{F}$ scales according to the SNL. \par 
Mixing the squeezed state with a coherent probe modifies the QFI and allows beyond-SNL scaling. As shown in Eq. (\ref{eq:SNR_photon_counting}) of the main text, photon counting after a Mach-Zehnder interferometer achieves an SNR scaling of $N^{3/2}$; note that this result is true despite $\braket{(\Delta x)^2},\braket{n_s}$ scaling as $N^{-0.5},N^{+0.5}$respectively. The conceptual shift from a phase space-based perspective to a measurement-aware analysis of Fisher information reveals the metrological advantage of squeezed states.  \par
\if In the parametric approximation, $\braket{\Delta x^2}=\exp[-\zeta]$, where $\zeta$ is related to the squeezing strength and the pump photon number $N$, and the SNR equals $N\exp[\zeta]$ \cite{mandelandwolf1995book}. Since $\zeta\propto N$, the parametric approximation actually predicts an exponential increase in the SNR since the pump is treated as an undepleted source of energy.\par\fi
\emph{Stochastic pump -- } The possibility of achieving beyond-SNL performance in Mach--Zehnder interferometry, despite the phase space effects associated with pump quantization, can be motivated semiclassically. Rather than treating the pump as a fully quantized mode, let us first consider a classical pump field with stochastic phase fluctuations. In the parametric approximation, the pump mode is replaced by

$a_p \rightarrow |\alpha| e^{i\beta},$
where $N = |\alpha|^2$
and $\beta$ denotes the pump phase. Solving the Heisenberg equations of motion for $a_s$ generated by the Hamiltonian in Eq.~(\ref{eq:Hamiltonian}) of the main text gives
\begin{align}
a_s(t)
=
a_s(0)\cosh\left(|\alpha|\kappa t\right)
-
e^{i\beta}a_s^\dagger(0)\sinh\left(|\alpha|\kappa t\right).
\label{eq:parametric_solution}
\end{align}
We now regard the pump phase $\beta$ as a classical random variable with prescribed variance
$\overline{\Delta \beta^2} \sim N^{-1}$,
as appropriate for the phase uncertainty of a coherent state, where the overline denotes classical ensemble averaging. In phase space, these pump-phase fluctuations lead to a jitter of the squeezing ellipse orientation. As a result, even though each realization produces a squeezed state, averaging over the fluctuating pump phase degrades the observable squeezing. The minimum achievable quadrature variance is then limited to the scaling
$\langle \Delta x^2 \rangle_{\mathrm{min}} \sim N^{-1/2}$
~\cite{crouch_limitations_1988,kinsler_limits_1993}.

\par
While the phase space jitter seems to be an intrinsic limitation due to pump phase fluctuations, its effect can be suppressed by measuring in the frame that co-rotates with the squeezing ellipse. Operationally, this is precisely what homodyne detection achieves by mixing the squeezed state with a phase-locked coherent probe. To see this mathematically, consider the following. Assuming the signal at $t=0$ is a vacuum mode, and using the expression in Eq.  ($\ref{eq:parametric_solution}$), we find $\braket{\Delta a^2_s}$ proportional to $\overline {e^{i\beta}}$. One would expect degraded squeezing due to a fluctuating $\beta$. However, the QFI for phase estimation as given by Eq. (\ref{eq:QFI}) has a $\beta$-dependence only in the last term containing the factor $e^{-i2\phi}\braket{\Delta a^{2}_s}$. For a stochastic pump, this term is proportional to $\overline{e^{-i(2\phi-\beta)}}$. Crucially, because the pump field is generated by second harmonic generation of the probe, as is necessary to maintain a phase-locked probe for homodyne detection, we have the (deterministic) relation $(2\phi=\beta)$ \cite{friberg_production_1984,afek_high-noon_2010} and $e^{-i2\phi}\braket{\Delta a^{2}_s}$ is invariant under pump phase fluctuations \footnote{We assume that the second harmonic generation does not add independent fluctuations}. Thus, the QFI does not degrade in the presence of a stochastic pump. Similar arguments apply to the SNR of photon counting as given by Eq. (\ref{eq:SNR_photon_counting}) of the main text, showing that any claim of a metrological advantage, or lack thereof, is meaningful only in relation to phase encoding and measurement protocols.\par
\emph{Energy Balance -- }Further modifications to the parametric approximation have also been considered  \cite{paris_small_1995,barnett2003ultimate}. In particular, Paris \cite{paris_small_1995} approached squeezed-state interferometry from an energy balance perspective. They showed that under a fixed total photon budget \textit{within the interferometer}, sub-shot-noise scaling of $\braket{\Delta^2\phi} \sim N^{-3/2}$ can be achieved, although this enhancement rapidly degrades with detector inefficiency. These works, however, treat the squeezed state as an external resource and do not account for the cost or dynamics of its generation from a quantized pump.
\if Beyond the undepleted-pump approximation, several works have examined how pump quantization and depletion influence the generation of squeezed light and its metrological utility \cite{crouch_limitations_1988,kinsler_limits_1993,kinsler_limits_2001}. These works  numerically considered the case when the squeezer is biased at the point of maximum squeezing $t_{sq}$. At this point, it was shown that the squeezed quadrature variance $\braket{\Delta x^2}$ scales as $N^{-1/2}$, while the mean flux of the squeezed light scales as $N^{1/2}$. This can be interpreted as a direct consequence of pump quantization: the dynamical phase fluctuations of the pump limit the available squeezing, and the finite pump energy limits the generated flux. Consequently, the SNR, given by $\braket{n}_s/\braket{\Delta x^2}$, scales as $N$ which is just the SNL.\par
Both the , when the  analyzed the quantum nature of the pump by introducing a phase-resolution metric for the generated state, arguing that the intrinsic phase uncertainty of the pump determines the squeezing axis and ultimately bounds the achievable phase sensitivity. In their analysis, the degree of squeezing scales as $N^{-1/2}$ with the number of pump photons $N$, implying that the phase resolution of any interferometric scheme employing the resulting squeezed vacuum cannot surpass this limit. This interpretation suggests that the advantage of squeezing lies primarily in reduced power requirements rather than in improved scaling. However, their treatment does not explicitly consider the dynamical effects of pump depletion or the emergence of non-Gaussian correlations. In contrast, Paris \emph{et al.} \cite{paris_small_1995} approached the problem from a metrological perspective, using error-propagation analyses of interferometers fed with a coherent probe and a squeezed vacuum. They showed that under a fixed total photon budget within the interferometer, sub-shot-noise scaling of $\Delta\phi \sim N^{-3/4}$ can be achieved, although this enhancement rapidly degrades with detector inefficiency. These works, however, treat the squeezed state as an external resource and do not account for the cost or dynamics of its generation from a finite pump. Further studies by Ma\^{i}tre \emph{et al.} have modeled pump depletion beyond the simple parametric approximation, revealing that strong coupling drives the evolution of non-Gaussian, entangled pump–signal–idler states. Yet these analyses focus primarily on state properties—such as squeezing, entanglement, and photon statistics—rather than on how these features translate into phase-sensitivity bounds or optimal measurement strategies \cite{florez_pump_2020}. Consequently, a unified framework connecting pump quantization, state generation, and metrological performance remains lacking. Our work addresses this gap by explicitly treating the pump as a quantized, depleting phase reference and analyzing how its dynamics modify both the achievable sensitivity and the measurements required to attain it.
\fi
\section*{Appendix C: QFI Scaling for different interaction lengths}
\begin{figure}[t!]
    \centering
    \hspace*{-1cm} 
    \includegraphics[trim={0cm 0cm 0cm 0cm},clip,scale=1]{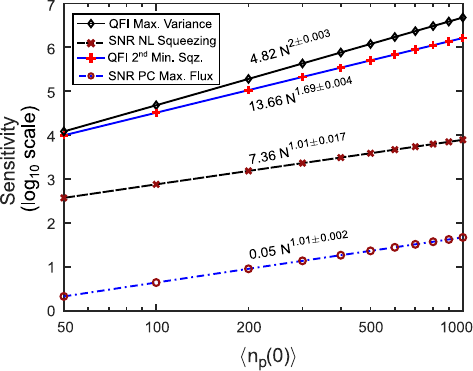}
    \caption{\textbf{QFI Scaling for the DPO at different interaction lengths}. For the DPO biased at maximum signal photon variance ($t_{var}$), the QFI (solid black line and black diamonds) shows Heisenberg-limited (HL) scaling. For the DPO biased at the second maximum of squeezing (second minimum of $\braket{\Delta x^2}$) ($t_{sq2}$), the QFI (solid blue line and red plus signs) does not attain the HL, but instead scales as $N^{1.69\pm0.004}$, still beating the shot noise limit (SNL) and the performance at the first squeezing maximum. For the nonlinear measurement in Fig. (\ref{fig:4}) (dashed black lines and brown crosses) operating at maximum squeezing $t_{sq}$ the sensitivity shows SNL scaling, while the Mach-Zehnder scheme shown in Fig. (1) achieves the same SNL scaling when biased at maximum signal flux $t_{fl}$. Note that the dashed and solid lines are formed by connecting the data points computed numerically, while the scaling shown is obtained from a linear fit to the numerical data. Table II lists the different operating points for these curves.}
    \label{fig:5}
\end{figure}
In this section we list the prefactors of the scaling curves shown in Fig. (\ref{fig:3}). We also show the scaling for QFI, photon counting SNR $(S_{PC})$, and the SNR for nonlinear (NL) measurement $S_{NL}$ at points other than those corresponding to maximum squeezing and maximum flux. \par 
\begin{table}[t]
\centering
\caption{Prefactors for scaling curves shown in Fig. (\ref{fig:3}) of the main text. QFI is given by Eq. (\ref{eq:QFI}), the photon counting SNR $(S_{PC})$ is given by Eq. (\ref{eq:SNR_photon_counting}), while the SNR for the nonlinear measurement $(S_{NL})$ is computed numerically.}
\label{tab:scalings}
\begin{tabular}{lllll}
\toprule
Observable &
Interaction time &
Scaling \\
\midrule

QFI  &
$t_{fl}$ (max. flux) &
$5.53N^{2\pm0.00024}$ &
 \\

$S_{\mathrm{NL}}$ &
$t_{fl}$ &
$1.03\,N^{2\pm0.004}$ &
 \\

QFI &
$t_{sq}$ (max. squeezing) &
$4.68N^{1.48\pm0.008}$ &
 \\

$S_{\mathrm{PC}}$ &
$t_{sq}$ &
$0.56N^{1.52\pm0.012}$ &
 \\

\bottomrule
\end{tabular}
\end{table}
Different interaction lengths and associated choice of measurements give a huge design space to explore for the DPO operating point. Table $\ref{tab:scalings}$ shows the prefactors for the scaling curves shown in Fig. (\ref{fig:3}) of the main text. Table II summarizes the other scalings explored in this Appendix.
\begin{table}[t]
\centering
\caption{Scaling shown in Fig. (\ref{fig:5}).}
\label{tab:scalings2}
\begin{tabular}{lllll}
\toprule
Observable &
Interaction time &
Scaling \\
\midrule

QFI  &
$t_{var}$ (max. variance) &
$4.82N^{2\pm0.003}$ &
 \\

$S_{\mathrm{NL}}$ &
$t_{sq}$ (max. squeezing)&
$7.36\,N^{1.01\pm0.017}$ &
 \\

QFI &
$t_{sq2}$ ($2^{nd}$max. squeezing) &
$13.66N^{1.69\pm0.004}$ &
 \\

$S_{\mathrm{PC}}$ &
$t_{fl}$ (max. flux)&
$0.05N^{1.01\pm0.002}$ &
 \\

\bottomrule
\end{tabular}
\end{table}
We find that the nonlinear (NL) scheme can give below-HL performance when operated at points other than $t_{fl}$. As shown in Table \ref{tab:scalings2} and Fig. (\ref{fig:5}), the NL scheme achieves SNL when operated at $t_{sq}$. Note that the QFI for the NL scheme at $t_{sq}$ shows shot noise scaling (See Fig. (\ref{fig:3}) in main text). In that sense, the NL scheme is optimal, but does not provide any quantum advantage. Conversely, the photon counting SNR $S_{PC}$ cannot go beyond the SNL when the system is flux-optimized at $t_{fl}$. Although both schemes operate at the SNL and are suboptimal, note that the prefactor for $S_{PC}$ is an order of magnitude smaller than the corresponding prefactor for $S_{NL}$ in Table \ref{tab:scalings2}. Of course, we are not restricted to $t_{sq},t_{fl}$. Figure (\ref{fig:5}) shows the QFI at the point of maximum signal photon number variance $t_{var}$, where the Heisenberg scaling persists. Notably, $t_{var}$ does not coincide with $t_{fl}$, as is also shown in Fig. (\ref{fig:2}c) for $N=100$. Nevertheless, as discussed in Appendix E, we expect the signal photon number variance at $t_{var}$ to scale as $N^2$, which matches the simulated scaling shown in Fig. (\ref{fig:5}). Similarly, instead of targeting maximum squeezing (minimum $\braket{\Delta x^2}$), one can bias at the second minimum of $\braket{\Delta x^2}$ (corresponding to second maximum of squeezing), an instance of which is shown around $\kappa t=0.6$ in Fig. (\ref{fig:2}b) for $N=100$. The QFI at this point scales as $N^{1.69\pm0.004}$, which is better than the $N^{3/2}$ scaling achieved by biasing at $t_{sq}$. This is expected since the signal photon number variance is higher at the second minimum of $\braket{\Delta x^2}$ than at the first occurring at $t_{sq}$. At the second minimum of $\braket{\Delta x^2}$, $S_{PC}$ (not shown in Fig. (\ref{fig:5})), shows a scaling of $0.01N^{1.64\pm0.06}$. Although this still beats both the SNL and $N^{3/2}$ scaling, it has a very small prefactor of $0.01$ and the actual value of the SNR, at least for the pump powers considered here, is smaller than $S_{PC}$ operating at $t_{sq}$. Note that we have restricted our analysis to the first maximum of signal photon mean and variance, and leave the analysis of long-term dynamics to future work.\par
\section*{Appendix D: QFI for phase encoding with photon number difference}
In this section, we compare the QFI scaling for a phase encoding in the Mach-Zehnder photon number difference operator, which is a different phase encoding scheme used in earlier works on quantum-enhanced phase metrology \cite{dowlingtakeoka2017fundamentalprecisionofMZI,lang_optimal_2013}. In the main text, we assumed that the measured phase is inserted in one arm of the MZI, with the optical path length $\theta$ being imprinted via the interaction Hamiltonian $e^{i\hat{n}\theta}$, with $\hat{n}$ being the number operator in the corresponding mode of the MZI arm. Let $n_{1,2}$ denote the photon number in the two arms of the interferometer. Strictly speaking, both arms impart a phase difference $\theta_{1,2}$, making phase measurement a two-parameter estimation problem. It is instructive to reparametrize the phase encoding interaction as $e^{i{(n_1\theta_1+n_2\theta_2})}=e^{in_+(\theta_1+\theta_2)/2}e^{in_-(\theta_1-\theta_2)/2}$, where $n_{\pm}=n_1\pm n_2$. For the cases where the optical path length in both arms needs to be estimated, such as in LIGO, the $n_+$ term cannot be ignored, and the metrological performance is limited by the multiparameter QFI. In this work, we consider the case corresponding to placing the phase object in one arm of the interferometer, while the global path lengths in the interferometer are known. The phase encoding can then simply be written as $e^{in_-\theta/2}$, where the phase difference $\theta=(\theta_1-\theta_2)$. The QFI for phase estimation is found as
\begin{align}
   \mathcal{F_{D}} &= \braket{\Delta n^2_-}\nonumber\\
   &=\braket{n_s}+|\alpha|^2\left(\braket{\Delta a^{\dagger}_s \Delta
  a_s}+\braket{\Delta
  a_s\Delta a^{\dagger}_s}\right) \nonumber\\
  &-2\Re{|\alpha|^2e^{-i2\phi}\braket{\Delta a^{2}_s}},
  \label{eq:QFI_MI}
\end{align}
where the subscript $D$ denotes using the phase difference for the phase encoding. Because we are using $n_-$ to encode the phase, Eq. (\ref{eq:QFI_MI}) does not have the signal and probe variance terms that are present in Eq. (\ref{eq:QFI}), and the actual values predicted by the two equations will differ, as was also elaborated in \cite{dowlingtakeoka2017fundamentalprecisionofMZI}. Nevertheless, the second term in Eq. (\ref{eq:QFI_MI}) is equivalent to $N\braket{n_s}$, which scales as $N^2$ when the signal is flux optimized and $\braket{n_s}\propto N$. We find both $\mathcal{F_D}$ and $\mathcal{F}$, the QFI  considered in the main text, to have similar scalings and comparable prefactors when biased at $t_{sq},t_{fl},$ and $t_{var}$. Therefore, insofar as scaling is concerned, both encodings have similar performance for the DPO.
\section*{Appendix E: Evolution of the photon number distribution}
\begin{figure*}[b!]
    \centering
    \hspace*{-1cm} 
    \includegraphics[trim={0cm 0cm 0cm 0cm},clip,scale=1.45]{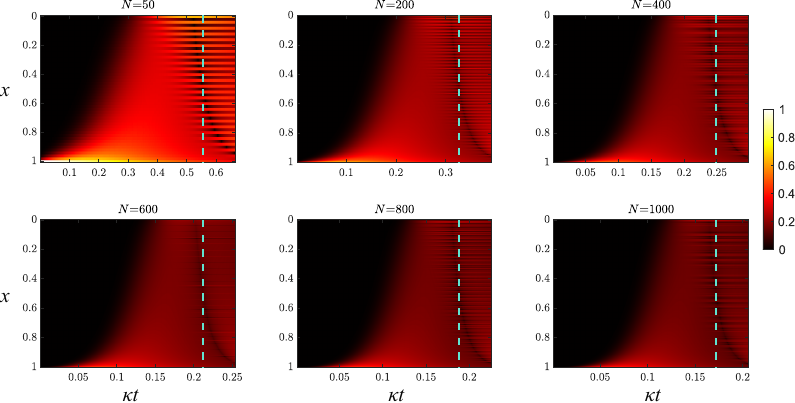}
    \caption{\textbf{Photon number evolution $c_n(t)$ for a Fock state pump}: The input state at $t=0$ is $\ket{0}_s\ket{N}_p$, which has a unit amplitude at $x=1$, where $x$ is the normalized population coordinate $x=n/N$. For enhanced contrast this figure plots $\sqrt{|c_{n}(t)|}$. The probability distribution expands in the Fock space, and the signal mean and variance grow until the full support of the Fock space is occupied. The cyan dashed line denotes the time $t_{fl}$ when the signal flux is maximized. For this figure, $\kappa=1$.}
    \label{fig:6}
\end{figure*}
In this section, we show the evolution of the photon number distribution for different pump photon numbers in a Fock state. In particular, we show that when the signal flux is maximized, the photon number distribution spans the full support of the signal Fock space. Qualitatively, this can be interpreted as one of the mechanisms enabling Heisenberg-limited performance of the joint signal-pump state. \par
Consider the input state $\ket{\psi(t=0)}=\ket{0}_s\ket{N}_p$, where the pump is in a Fock state $\ket{N}_p$. Evolution under the Hamiltonian $H_I$ given by Eq. (\ref{eq:Hamiltonian}) preserves the total input photon number N, and the time-evolved state is given by $\ket{\psi(t)}=\sum_{n=0}^{N}c_{n}(t)\ket{2(N-n)}\ket{n}$  with $c_{n}(0)=\delta_{n,N}$, and $|c_{n}(t)|^2$ describing the (discrete) probability distribution to observe $n$ pump and $2(N-n)$ signal photons. Figure (\ref{fig:6}) shows the evolution of $c_n(t)$ for Fock pumps with $N=[50,200,400,600,800,1000]$. Note that there are no odd signal photons in the evolved state and the odd indices are not indicated in the figure. To facilitate comparison between different $N$ values, we plot using the normalized coordinate $x=n/N$, where $x=0(1)$ denotes the amplitude for all $N$ photons to be present in the signal (pump) mode. The evolution is qualitatively similar across the different $N$ values. The initial energy depletes in the pump mode ($x=1$), and the photon population spreads across the Fock space. Once the probability amplitude reaches the $x=0$ boundary, it undergoes a reflection and we see oscillations in the probability distribution \cite{schleich1987oscillations}. These oscillations of probability distribution are qualitatively similar to quantum caustics observed in many-body photonic, spin, and condensed matter systems governed by nonlinear Hamiltonians \cite{shchesnovich2008fock,kirkby2019quantum}. Strictly speaking, the probability distribution is discrete, but for large $N$ the sampling in the domain $x=[0,1]$ increases, and the finer scale of oscillations might appear aliased in the figure. The range of $\kappa t$ in each plot in Fig. (\ref{fig:6}) is restricted to $1.2t_{fl}$ (for $\kappa=1$), where $t_{fl}$ is the time at which the signal flux is maximized, shown by the vertical dashed cyan line. As expected, $t_{sq}$ decreases with increasing photon number.\par 
The Quantum Fisher Information (QFI) for phase estimation as given by Eq. (\ref{eq:QFI}) has one term proportional to the signal photon variance $\braket{\Delta n_s^2}$, and one proportional to $N\braket{n_s}$. Figure (\ref{fig:6}) shows that at $t_{fl}$, the wavefunction spans the full support of the Fock space from $n=0$ to $n=N$. Therefore, the signal mean and variance are expected to scale as $N,N^2$ respectively. Consequently, the QFI will scale as $N^2$.  Note that Fig. (\ref{fig:6}) assumes a Fock state pump. To compute the variance and mean for a coherent state pump considered in the main text, one has to average over the pump Poisson distribution. However, each pump Fock state will evolve into a broad signal distribution spanning the full support of the Hilbert space. The QFI for a coherent state pump is then also expected to scale as $N^2$, as shown in Fig. (3) of the main text.\par

\end{document}